# BioinspiredLLM: Conversational Large Language Model for the Mechanics of Biological and Bio-inspired Materials

*Rachel K. Luu, Markus J. Buehler\**

R. K. Luu, M. J. Buehler
Laboratory for Atomistic and Molecular Mechanics (LAMM), Massachusetts Institute of Technology, 77 Massachusetts Avenue, Cambridge, MA, 02139, USA

R. K. Luu
Department of Materials Science and Engineering, Massachusetts Institute of Technology, 77 Massachusetts Avenue, Cambridge, MA 02139, USA

M.J. Buehler
Center for Computational Science and Engineering, Schwarzman College of Computing, Massachusetts Institute of Technology, 77 Massachusetts Avenue, Cambridge, MA 02139, USA

\*E-mail: mbuehler@mit.edu

**Abstract:** The study of biological materials and bio-inspired materials science is well established; however, surprisingly little knowledge has been systematically translated to engineering solutions. To accelerate discovery and guide insights, an open-source autoregressive transformer large language model (LLM), BioinspiredLLM, is reported. The model was finetuned with a corpus of over a thousand peer-reviewed articles in the field of structural biological and bio-inspired materials and can be prompted to recall information, assist with research tasks, and function as an engine for creativity. The model has proven that it is able to accurately recall information about biological materials and is further enhanced with enhanced reasoning ability, as well as with retrieval-augmented generation to incorporate new data during generation that can also help to traceback sources, update the knowledge base, and connect knowledge domains. BioinspiredLLM also has been shown to develop sound hypotheses regarding biological materials design and remarkably so for materials that have never been explicitly studied before. Lastly, the model showed impressive promise in collaborating with other generative artificial intelligence models in a workflow that can reshape the traditional materials design process. This collaborative generative artificial intelligence method can stimulate and enhance bio-inspired materials design workflows. Biological materials are at a critical intersection of multiple scientific fields and models like BioinspiredLLM help to connect knowledge domains.

**Table of Contents Entry:** BioinspiredLLM, a finetuned conversational large language model specializing in biological material mechanics, can assist with accelerating research through guiding insights, new hypotheses, and collaborative generative artificial intelligence efforts that can shift paradigms in materials design and the production of bio-inspired materials. The field of biological materials is incredibly interdisciplinary and models like BioinspiredLLM help to bridge knowledge domains.

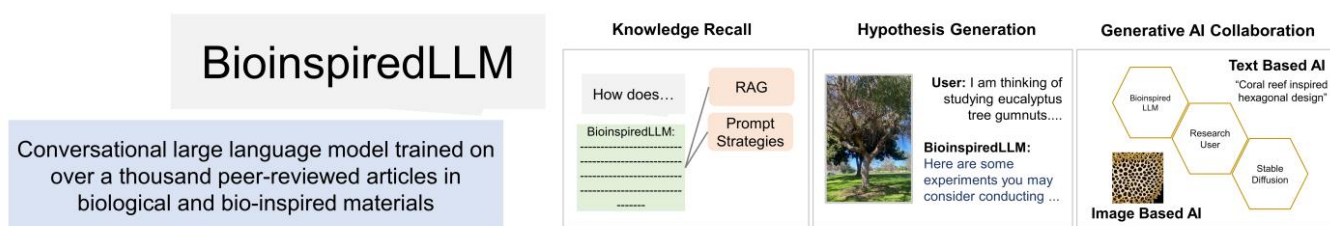

# 1. Introduction

Biological and bio-inspired materials science has been studied for generations of materials scientists, biologists, and engineers, due to its impressive hierarchical structure-property relationships [1–8] and opportunity to provide blueprints for creating sustainable, high-performance materials [9–16]. Countless multiscale studies have been performed on biological materials such as armadillo shells[17], bamboo[18], coconut husks[19], and sheep horns[20] to only name a few. However, much of the knowledge in the field is characterization focused and many biological subjects have not yet reached bio-inspired engineering scale applications. With ongoing environmental challenges, there is a strong urge to learn how to develop sustainable materials systems from Nature [21–25].

With the growth of materials informatics tools, there is an opportunity to accelerate discovery and development of bio-inspired solutions [26]. One tool being large language models (LLMs), which are generative Artificial Intelligence (AI) models trained on an extremely large amount of text known as a corpus, usually of general information from articles and websites. There has been strong inclination to finetune pretrained language models so that the models become specialized in scientific topics such as battery research: BatteryBERT [27], optical materials: OpticalBERT [28], proteins: ProGen [29], and polymers: TransPolymer [30]. These finetuned models have shown to outperform their original models in their specialized tasks. For the case of BatteryBERT and OpticalBERT, BERT is a family of models that are bidirectional transformers that use encoder self-attention layers to build contextualized representations that are helpful for natural language processing tasks such as sentiment analysis and entity recognition but make them less suitable for developing dynamic interactive dialogue systems. As for the other two models, the architectures are also focused on generating sequences and predicting properties and are also less for conversational outputs. On the other hand, LLM architectures such as ChatGPT and Bard have grown in popularity due to their autoregressive decoders that allow for the ability to hold conversations over thousands, and in some cases, tens of thousands of words [31–33]. Now, with the emergence of open-source autoregressive decoder frameworks, the opportunity to explore specializations in conversational LLMs is possible.

The study of mechanics of biological materials poses as a unique opportunity for finetuning an LLM for reasons such as there being a large wealth of research in the field which can supplement the training dataset, and the need to relate, translate and connect between knowledge domains. There is also likely a strong baseline familiarity of biological materials in the pretraining, where the average knowledge of nature, animals, and plants is most likely greater than, say, chemical compounds or protein sequences. There is also no set standardization in discussing or describing complex, hierarchical materials. The way that molecules have Simplified Molecular Input Line Entry System (SMILES) representations, natural biological materials lack clear formats of written representations but have been described in a multitude of ways in the literature. Therefore, the subject of biological materials and mechanics is an ideal candidate to gain insights about how an autoregressive large language model could synthesize literature; and, with that synthesis, be able to assist with the materials discovery and design process. Specializing a LLM with the fundamental knowledge of biological materials mechanics can serve as a step in the right direction in terms of accelerating research and scientific discovery in bio-inspired materials.

## 2. Results and Discussion
### 2.1. Model Training and Evaluation Strategy

Llama-2-13b-chat, an open-source conversational large language model, was selected as the base model. Orca-2-13b[34] is a finetuned version of Llama-2-13b-chat[35] that has enhanced reasoning abilities and was then fine-tuned using our final corpus dataset. Composed of over a thousand articles in the field of structural biological materials, **Figure 1a** displays the breakdown of the finalized corpus by publisher and publishing year, color indicating the most frequently occurring publishers: Elsevier (light blue), Wiley (yellow), Springer Nature (dark green), and American Chemical Society (dark blue). The corpus of article full texts was extracted and cleaned in the procedure outlined in **Section 4, Methods**. With the clean text of the articles, two approaches were conducted for the development of the training set. The first approach was with using text as is, generating token lengths as default. The second approach, we call Q-A Processing, uses the original non-fine-tuned Llama-2-13-chat model to process and clean the text content even further and to distill key insights in question-answer pairs. This was done by firstly parsing the text into chunks. These chunks are then queried to Llama-2-13b-chat (in principle, any general-purpose LLM can be used for distillation), asking for a question that can be answered by a given text chunk {txt}:

```
Give me a concise question to which the answer is "{txt}". Answer as a question, one sentence, short.
```

Then, after noting the 'question' {question}, Llama-2-13b-chat is asked to rephrase the same chunk of text to fine comb and remove any artifacts not removed in the first round of cleaning:

```
Write a succinct summary of key concepts of how "{txt}" answers "{question}". The summary must stand on its own. Never include math, equations, variables, and numbers in the response.
```

The response provided is then noted as the 'answer'. The full Q-A processed dataset and the list of all articles featured in the development of the dataset are provided in the Supporting Information files.

After finetuning the model independently with both approaches, it was apparent that the Non Q-A default method model tended to produce responses with more undesirable remnants. **Figure 1b** shows an example of the non Q-A model echoing the mannerisms and language from the raw data such as saying "In our previous study," or referencing nonexistent figures or videos. After applying the Q-A processing method on the dataset, where the article text goes through another layer of processing, it is apparent the token length shows a more aligned distribution, as shown in **Figure 1c**. **Figure 1d** shows an interactive Gradio chat interface of a Q-A processed model with an example query and response, now rid of any undesirable remnants. Upon inspection of responses by the two approaches, there does not appear to be a large discrepancy content-wise between the non Q-A dataset and the Q-A processed dataset approach. Since the Q-A processed model performs better conversationally, all experiments here forth are using the Q-A processed approach. We finetune both Llama-2-13b-chat and Orca-2-13b independently though find better performance with the Orca-2-13b model and henceforth will refer to the Orca-2-13b finetuned version as the main model: BioinspiredLLM. For comparison purposes, the finetuned Llama-2-13b-chat model will be referred to as Llama-BioLLM. We refer to the original, non-finetuned models as Llama 13b-chat and Orca 13b from here forth.

**Figure 2** depicts the final framework of BioinspiredLLM architecture, including the autoregressive transformer model and a flowchart of how BioinspiredLLM functions on graph-forming attention mechanisms from system prompts to queries to its generated responses.

At the time of writing this paper, the family of Llama 2 models is one of the top open-source LLMs determined by automated evaluation on huggingface.co (https://huggingface.co/spaces/HuggingFaceH4/open_llm_leaderboard) where LLMs are tested with few-shot and zero-shot evaluation benchmarks that focus on reasoning and general knowledge from a wide variety of fields including science, math, history, and law. Therefore, within the scope of this work, the focus is on manual evaluation and comparison between models. Queries to assess BioinspiredLLM were carefully curated to assess model capabilities loosely following the framework outlined by Brodnik et al. [36] which include three main tasks 1. Knowledge Recall: rapidly provide information about a large span of documented biological materials, 2. Hypothesis Generation: generate insights about biological materials research including experimentation, subject selection, and applications, and lastly 3. Assistive Tasks: provide help to researchers in isolated tasks including but not limited to prompt engineering and clustering.

## 2.2. Knowledge Recall

When prompted, BioinspiredLLM can be used to rapidly recall information about biological materials. BioinspiredLLM, Llama-BioLLM, and their original base models: Llama 13b-chat and Orca 13b were evaluated using the set of challenging biological materials exam questions and inference parameters described in **4. Methods.** This exam was purposefully composed to be a very challenging task of zero-shot generation that is meant to rigorously stress-test the knowledge recall abilities of the models. Additionally, to account for worst case scenario inferencing, no system prompts were set. The questions were categorized by General (broad, general questions about biological materials), Specific (specific questions about particular biological material species), Non-Biological (compares synthetic materials to biological materials), and lastly a small portion of the questions are Numerical (recalling specific number values for mechanical properties or structures). The resulting model total scores are displayed in **Figure 3a** and a breakdown of categories and performance are shown in **Figure 3b**. BioinspiredLLM (with and without Retrieval-Augmented Generation) outperforms the rest of the models in all categories. Llama-BioLLM outperforms the base models, however, does not perform as well as the Orca finetuned BioinspiredLLM most likely due to the enhanced reasoning abilities that were trained into the Orca version.

Unique to the model architecture focused on enhanced reasoning capabilities, we show that we can further supplement BioinspiredLLM's score by making minor adjustments when prompting the question and/or adding a system prompt. Taking a question that BioinspiredLLM initially answered incorrectly and adding "Think step by step:" to the input prompt, we find that the correct answer can be achieved for a portion of the initially incorrect questions. When applying this method, BioinspiredLLM's baseline score increases from 82% to a performance of about 90% on the biological materials exam questions.

Despite already achieving solid performance, to further supplement knowledge recall scenarios retrieval-augmented generation (RAG) methods[37,38] can be integrated to provide BioinspiredLLM an external source of knowledge to formulate answers (the retrieval system is set up so that it can find relevant documents from a database or corpus of data; this database can consist of a wide range of text sources and other materials, see **Materials and Methods** for details). When doing so, BioinspiredLLM with RAG methods claims the top score in the examination among all categories. Particularly so, we find that RAG can be employed to support BioinspiredLLM in numerical knowledge recall, a category of questions that all other models underperformed in. We prepared a vector database based on the original the training dataset of scientific articles. Subsequently, when an RAG-specific query is made, the database is systematically scanned using vector embeddings to retrieve relevant information for the query. BioinspiredLLM is then supplemented with that information before the model generates an answer. A framework of this process is shown in **Figure 3c**. An example for a numerical question and non-numerical question is presented wherein a previously inconsistently answered question now yields the correct response. Importantly, this method also offers a mechanism to trace back the precise source of information used to generate the answer, in this case it traces back to articles by Lauer *et al.*[39], Naleway *et al.*[40], and Zhang *et al.*[41] Nonetheless, this biological materials examination experiment stresses the importance of utilizing available tools. Like with other modeling techniques, it is suggested that the best strategies and advanced tools are employed to increase performance and validate results.

Furthermore, highlighting the ingenuity of RAG methods, RAG can also be employed to maintain the model with up-to-date information in the database without needing to entirely re-train the LLM. In the following example, an article on further elucidating the study of equine hoof wall porosity, published after the corpus collection date, was collected into the RAG database for BioinspiredLLM to refer to [42]. In **Figure 4a**, a sample query is shown asking about this article and the generated response from BioinspiredLLM shows its ability to contextualize and summarize this new article that was not initially in the training set. Additionally, by using RAG methods, we can introduce articles from other knowledge domains, outside of biological material mechanics and use the model to integrate and connect knowledge, ideas and concepts. In **Figure 4b**, we add an article from the field of paleontology and ecology that touches on the evolutionary history of hooved mammals in the North American grassland biome[43]. We also added an article from the field of manufacturing that is a review article of rapid prototyping techniques[44] to the database. A sample query is shown for both articles, drawing connections between the content of the articles and BioinspiredLLM's finetuned knowledge of biological materials.

In another experiment, both BioinspiredLLM and Llama 13b-chat were prompted to produce their own multiple-choice questions and answers regarding biological materials:

```
Create a multiple-choice question about the mechanical behavior of a specific natural
biological material. Also, tell me the correct answer.
```

In this task, the base models Llama 13b-chat and Orca 13b particularly struggled, providing mostly incorrect answers and writing questions that included non-biological material options. BioinspiredLLM was able to successfully complete the task with mostly accurate questions and answers. Samples of the generated questions are shown in **Figure 5a** where the highlighted answer option is the suggested answer from the model, where incorrect answers are highlighted red, correct answers are highlighted green, and one ambiguous grey sample from BioinspiredLLM is also shown. In that case, BioinspiredLLM's selected answer suggests that turtle shells have a hollow structure that absorbs energy upon impact. Debatably, the turtle shell structure would be described more as "porous" than hollow, despite this imperfect answer, an explanation was further provided in BioinspiredLLM's generated responses that could allow for further elaboration. It can be observed that the questions generated by the base models tended to be vaguer in broadly comparing biological materials such as silk vs. bamboo vs. bone. Whereas, for BioinspiredLLM, the questions provided were detailed and focused on the mechanics of a singular specific biological material.

Given BioinspiredLLM's stronger accuracy, the model also proves beneficial in synthesizing information, avoiding the need to parse through the literature. A simple test was done using the Google search engine, Llama 13b-chat, Orca 13b, and BioinspiredLLM, querying each one about a specific biological materials question, in this case:

```
How does water content affect horse hoof wall mechanics?
```

**Figure 5b** shows the comparison of the four responses. In traditional web searching, the results provided are mostly links to leading articles in the field, many of which are included in the dataset. However, the traditional search does not immediately provide an answer and further manipulation is needed. Llama 13b-chat provided a long response that initially was accurate in describing how hydration affects mechanics but the response traverses off-topic into more of a veterinary perspective: "A balanced diet and regular grooming can help maintain healthy hoof moisture levels". A similar case was found for Orca 13b, where the response was more on-topic but still does not clearly state the mechanical trend. On the other hand, BioinspiredLLM provides a succinct response reporting the connection between hydrated hoof being more pliable and drier hoof being more prone to brittle fracture [45–47]. Since BioinspiredLLM is trained specifically on mechanics of biological materials, the model does not require as much context to provide clear and concise answers.

In another test usage, the models were asked for a list of biological materials that fit a criterion: being impact resistant. The models were queried five times and the instances of terms that appeared were counted. Top terms recorded are displayed in **Figure 6** for BioinspiredLLM and the base models. In Llama 13b-chat's produced responses, there are many off-topic materials that would not be considered biological such as Kevlar and titanium. In Orca 13b, there were only four materials mentioned in total and each was simply naming biological compositions and not specific biological species. At the top of the list for BioinspiredLLM are crustacean exoskeletons which include the specific mentions of lobster, crayfish, and crab shells. The biological species listed by BioinspiredLLM are firmly reported as "impact resistant" by studies [48,49].

**2.4. Hypothesis Generation**

BioinspiredLLM can also serve as an engine for scientific creativity. The model can answer open-ended questions about subjects it has never seen before and propose new predictions or hypotheses to help guide researchers. For example, users can ask the model about experimenting with subjects that have not been expressly studied before in the literature.

```
User: I am thinking of studying eucalyptus gumnuts for structure and mechanical properties.
Name some experiments I should do.

BioinspiredLLM: To study the structure and mechanical properties of eucalyptus gumnuts, you may consider the following experiments:
1. Material characterization: Measure the density, moisture content, and thermal conductivity of the gumnuts to understand their physical properties.
2. Microstructure analysis: Examine the gumnuts under a scanning electron microscope (SEM) to observe their microstructural features, such as cell walls, pores, and fibers.
3. Mechanical testing: Perform uniaxial tension tests on the gumnuts to determine their mechanical properties, such as Young's modulus, tensile strength, and elasticity.
4. Fracture analysis: Study the fracture surfaces of the gumnuts using SEM to understand the mechanisms of failure and the role of microstructural features in determining their mechanical properties.
5. Thermal analysis: Measure the thermal conductivity and specific heat capacity of the gumnuts to understand their thermal properties.
6. Water absorption and swelling: Measure the water absorption and swelling behavior of the gumnuts to understand their hydrophilic properties.
7. Biodegradation: Study the biodegradation of the gumnuts over time to understand their environmental impact and potential applications.
These experiments will provide valuable insights into the structure and mechanical properties of eucalyptus gumnuts, which can be used to develop new materials and technologies inspired by nature.
```

In this example, the user asks about studying eucalyptus gumnuts, which are small woody vessels that grow on eucalyptus trees, which by search appears to not have been explicitly studied in literature before. Here, BioinspiredLLM draws on its general knowledge from its pretraining and synthesizes it with the knowledge of biological materials and how to characterize biological materials from its finetuning dataset. BioinspiredLLM most likely relies on its pretraining to recognize what a eucalyptus gumnut is and then proposes experimental suggestions based off articles from its finetuning that concern characterizing plant material [50–58], notably including experiments on water uptake and biodegradation which both factors do typically significantly affect biological material mechanics.

In another experimental usage, BioinspiredLLM is asked for a hypothesis regarding the mechanical properties of jackfruit as shown in **Figure 7**. While jackfruit has been studied only just recently by Lazarus *et al.* [59], the work was not captured in this current dataset as it was published just outside of the dataset collection period. Therefore, BioinspiredLLM has never 'seen' this work before. A comparison of BioinspiredLLM's response (**Figure 7a**) and a figure drawn directly from Lazarus *et al.* (**Figure 7b**) is shown. When asked for its hypothesis about jackfruit thorn structure, BioinspiredLLM predicts that the thorns form a network that help with energy absorption and impact loading which was a theory strongly supported by the findings by Lazarus *et al*. Interestingly, BioinspiredLLM also notes an important caveat, that the thorns also help to control the crack propagation. This is exactly what is exhibited during the compact tension experiments in Lazarus *et al.* where the crack is seen propagating around thorns along the foamy matrix in a controlled manner. Fascinatingly, BioinspiredLLM has predicted both these findings that are now documented research, despite never having been explicitly trained on this data. We can imagine that these creative insights provided by BioinspiredLLM could help to guide researchers and generate entirely new ideas that may not be immediately thought of otherwise.

## 2.5. Assistive Tasks

BioinspiredLLM can help with isolated research tasks. In the following excerpt, BioinspiredLLM is asked to assemble a dataset.

```
User: I am assembling a dataset of names of biological materials, their structures, and their
mechanical properties. Could you list biological materials, its structures present, and its
mechanical properties? For example, material: spider silk, structure: webs, fibrils, fibrous
structure, mechanical properties: high tensile strength, lightweight.

BioinspiredLLM: Here is a list of specific biological material, their structures, and their
mechanical properties:
1.  Material: spider silk
       a.  structure: webs, fibrils, fibrous structure
       b.  mechanical properties: high tensile strength, lightweight
2.  Material: muscles
       a.   structure: elastic fibers, fibrous structure
       b.  mechanical properties: high tensile strength, elasticity
3.  Material: bone
       a.  structure: cortical and cancellous tissues, fibrous structure
       b.  mechanical properties: high compressive strength, stiffness
4.  Material: nacre
       a.  structure: layered structure
       b.  mechanical properties: high tensile strength, stiffness
5.  Material: hooves
       a.  structure: lamellar tissue
       b.  mechanical properties: high compressive strength, toughness
6.  Material: shell
       a.  structure: calcareous or aragonite
       b.  mechanical properties: high compressive strength, stiffness
7.  Material: feathers
       a.  structure: fibrous structure
       b.  mechanical properties: lightweight, high tensile strength
8.  Material: wings
       a.  structure: chitin-based structures
       b.  mechanical properties: lightweight, high tensile strength
9.  Material: plant stems
       a.  structure: vascular tissue
       b.  mechanical properties: high compressive strength, elasticity
10. Material: skin
       a.  structure: collagen-based structures
       b.  mechanical properties: elasticity
These are just some examples of the vast diversity of biological materials found in nature,
each with unique structures and mechanical properties that may vary in different environmental
conditions.
```

BioinspiredLLM provides a clear and organized chart of species and their respective structures and mechanical properties tailored to the user's request. We can imagine using this response and extended responses to rapidly generate and fill in a full dataset of biological materials. Additionally, this approach limits the potential subjective prejudice that would arise if a human were to choose and group species and properties.

In a more unconventional, though fascinating, scenario, BioinspiredLLM can assist with prompt engineering and collaborate with other generative AI models. Collaborating with text to image models like Stable Diffusion 2.0[60], BioinspiredLLM can generate prompts and assist the user in brainstorming ideas. **Figure 8** shows a potential workflow showing the collaboration between BioinspiredLLM, Stable Diffusion 2.0, and a user. The user asks BioinspiredLLM for a concise yet descriptive prompt for bio-inspired designs and BioinspiredLLM provides options for bio-inspired materials for algae, feathers, spider webs, and coral. The user responds and asks for more prompts, helping to guide direction of design. Selected BioinspiredLLM suggested prompts were inputted into a diffusion-based image synthesis AI tool, Stable Diffusion 2.0[61], and the resulting images are depicted in **Figure 8b**. One selected image for one of the coral bio-inspired designs is selected and processed into a 3D model using heat mapping (similar approach as in [62]). These generated 3D models can then be used for future simulation (coarse grain analysis, finite element analysis) or experiment (mechanically characterizing 3D printed samples). The data from those studies could inform BioinspiredLLM to work with the user to adjust prompts accordingly.

In another example, as shown in **Figure 9**, the user asks BioinspiredLLM for ideas of two biological materials structures to combine. BioinspiredLLM offers creative suggestions such as combining plant cell walls and animal hooves, sea sponge spicules and bone, and lotus leaf and butterfly wings. Not only does BioinspiredLLM provide biological species and fascinating combinations, but each response also outlines the logic of the material selection in the context of material properties and proposes a hypothesis about the behavior of the new design. These design ideas are also input into Stable Diffusion 2.0 to produce 2D images, selecting one to convert into a 3D model. It is clear that these generative AI frameworks can drastically accelerate the creation of bio-inspired designs and prototypes. By tapping into the generative 'creative' capacity of BioinspiredLLM, researchers can be guided by unique ideas that are backed by mechanical insights. With the assistance of generative AI techniques, the timeline for bio-inspired materials design and development can be remarkably expedited.

## 3. Conclusions

BioinspiredLLM is a fine-tuned conversational large language model that serves as an expert in structural biological materials mechanics and can help to accelerate research in the field. As main takeaways of this work, we find that:

- By using cutting edge techniques in deep learning, we have finetuned a conversational large language model to specialize in biological materials that considerably outperforms its base model.
- Our text/data mining and cleaning procedures in the Q-A processing distillation technique has shown to significantly reduce text fragments and provide realistic dialogue, despite being trained on a corpus of organized, formal writing.
- BioinspiredLLM has shown to excel in accurately and concisely recalling general information about biological materials especially when supported with retrieval augmentation strategies (RAG) that can not only provide additional context but also provides traceback to original references from which the data came from.
- BioinspiredLLM, via its abstraction and reasoning capabilities, generates interesting and creative insights about biological materials that have yet to be studied by synthesizing knowledge from its pretraining and finetuning.
- BioinspiredLLM can support researchers in dataset generation and other grouping or clustering like tasks.
- BioinspiredLLM has shown great promise in workflows that involve collaborating with other generative AI models. These new generative AI collaborative agent frameworks can significantly decrease the time and resources needed to design bio-inspired materials.

### 3.1. Limitations and Disclaimers

Like in all techniques of modeling, there are possibilities of errors. The original Llama 2 and Orca 2 models were strongly aligned to not spread misinformation and produce safer responses [63]. As a result, BioinspiredLLM has inherited these traits and performs reasonably well in these dimensions. However, it is still of utmost importance for researchers to also verify responses and avoid propagating errors, as discussed in recent literature[64] – a standard practice across all modeling techniques. To minimize risk of mistakes, employing chain-of-thought prompting and RAG methods, as introduced, proves beneficial. Additionally, the system prompt of BioinspiredLLM can be edited to guide context; we use, for instance:

```
You are BioinspiredLLM. You are knowledgeable in biological and bio-inspired materials and
provide accurate and qualitative insights about biological materials found in Nature. You are
a cautious assistant. You think step by step. You carefully follow instructions.
```

With the use of generative AI in science and other areas, it is important to consider authorship. For informational responses, users are urged to search the corpus dataset to find and recognize the original authors to be cited as sources or even easier if using RAG, trace back sources in the database. As for generative responses such as creative and original ideas that BioinspiredLLM provides, it is highly recommended to disclose the model's contribution with some level of disclaimer such as "BioinspiredLLM: a conversational large language model finetuned on a corpus of a thousand biological and bio-inspired material mechanics peer-reviewed articles." Even with original ideas, there should be thoughtful consideration of the fundamental groundwork laid by the authors' articles in the dataset and such authors should be rightfully cited as so.

Additionally, users should recognize the limitations of the corpus. In this version of BioinspiredLLM, over a thousand articles were included in the dataset and further details about its creation are outlined in section, **4. Methods**.

## 3.2. Outlook and future work

Language is arguably the most fundamental symbol-based form of communication, serving as the foundation for scientific and other forms of communication, mathematics, and engineering fields such as design. Models such as BioinspiredLLM contribute to the development of more powerful and versatile AI models that can help advance scientific research and solve complex problems in various applications. Moreover, as all models – be it experimental, computational (or analytical), they must be understood, used and applied in a setting in which they have been developed and validated for. As is generally true, their usefulness exists in the context of the question asked, its strength and weaknesses, and in a broader context of a particular research question. As tools of scientific inquiry, they must be considered as an ensemble of tools in the portfolio of scientists. LLMs are particularly intriguing due to their ability to reason and to apply knowledge recall, synthesis, and translation across multitudes of domains from logic to mathematics to simulation data [31,65–69].

There exist many further opportunities for research, including mechanisms to assess confidence levels of predictions, the development of active learning environments, and others. Another extension of this work would be fine-tuning a larger version of BioinspiredLLM based on the much larger 70b Llama-2 models that have been shown to offer a potentially better performance, albeit, at the cost of much great hardware requirements for both training and inference. There are other recent open-source developments, such as the Falcon open-source models with 180 billion parameters, or the Mistral series of models including new developments such as mixture-of-expert strategies that can be explored in future work.

Other improvements could be focused on the developments of larger and better curated datasets, complemented with human-derived question-answer pairs, and more detailed training including mathematical concepts, equations, tabulated data, figures, plots, and other aspects of critical knowledge. This can lead to more precise insights into biological materials, capturing the intricacies of structure and properties that are highly dependent on environmental and experimental conditions. Another area of great potential is the use of multimodal LLMs that can produce numerical output, code and field or image data such as the MeLM model [70]. In future studies, we are also looking into the difficult task of extracting more from journal articles including tables, equations, and figures which can bring more precise numerical data as well as content for both model generated plots as well as models helping user analyze their own data/plots.

In other possible future iterations of the model, enlarging the corpus is also an option. Of the 1,034 articles obtained for training this model, only 387 were some forms of open access designation, with non-open access articles proving more difficult to obtain through text/data mining. The development of new AI tools will benefit from openly accessible datasets, papers, where Open Access publishing and models such as FAIR are successful models [71]. Another future avenue can be utilizing multiple corpora from varying scientific disciplines to create a combined specialized model. For example, combining a biological materials science corpus with a manufacturing and fabrication corpus to create a model that could hypothetically be named 'BioinspiredFabLLM' with an emphasis in fabricating bio-inspired materials. Due to the intrinsic flexibility of the architecture such tasks can be easily accomplished and be based on the existing finetuned model via further training of the adaptor layers.

## 4. Materials and Methods

*Large Language Model Selection and Training*

Llama 2 is a set of high-performance open-access large language models released by Meta on 18 July 2023. The models are pretrained using an optimized autoregressive transformer using 2 trillion tokens of cleaned, publicly available online data. The models have a context length of 4,096 tokens and the "chat" models were developed through supervised fine-tuning and then reinforcement learning from human feedback to align the model with human preferences and to improve safety[63]. Orca 2, released by Microsoft, uses Llama 2 as a base model and is finetuned for enhanced reasoning abilities using synthetic data[34]. For developing BioinspiredLLM, Llama-2-13b-chat and Orca-2-13b was selected as the framework to be finetuned. In the naming convention, 13b (13 billion) indicates the number of parameters used to pretrain the model and chat indicates its optimization for dialogue applications. Due to noticeable increase in performance using the Orca model, we call that model the final finetuned model: BioinspiredLLM.

Using our generated dataset, we finetune the model using a Low-Rank Adaptation (LoRA) strategy[72,73] to ensure that the model learns the new tasks, but also avoids catastrophic forgetting by retaining the original knowledge. The LoRA strategy is a method that helps to generate adaptable and efficient LLMs as it speeds up the training while using less memory. Instead of retraining the entire model during fine-tuning, LoRA freezes the weights and biases of the pre-trained model and adds smaller trainable layers to each model layer. These layers help the model adapt to different tasks without changing all the parameters. We use a LoRA rank of $r = 96$ with $\alpha = 16$; models are quantized to 4 bit during training and inference to make their use accessible on relatively modest hardware (we use "nf4" quantization to convert data in floating point 32 bits (FP32) to a smaller precision, here integer 4 bits (int4) [73].

The models are developed in PyTorch [74] and implemented within the Hugging Face ecosystem (note the Llama 2 license that applies to all derivative works, as specified here: https://github.com/facebookresearch/llama/blob/main/LICENSE).
This architecture features 40 transformer layers and uses rotary positional embedding, which enables it to achieve long context lengths that can be extended easily via additional training.

We use a paged AdamW optimizer [75] with a learning rate of LR=0.0002 and $\varepsilon$ =1E8, and gradient norm clopping of 0.3. The Hugging Face Accelerate package (https://huggingface.co/docs/accelerate/index) is used to parallelize training. The model is trained for approximately 3,000 steps. The training objective used here is to maximize the likelihood of predicting the next token (*i.e.*, a letter, part of or a word) given the entirety of previous words (in a conversational model this context typically includes the system prompt to condition the behavior of the model, as well as all previous chat interactions between a user and BioinspiredLLM), for the training set consisting of text passages or the question-answer pairs. For each position in the sequence sample considered, the model estimates the probability distribution over the vocabulary for the next token, and the target is the actual next token.

*Dataset Generation*

For the scope of this work, the dataset is focused on biological material mechanics. The corpus selected was determined by using a search on Web of Science Core Collection (https://www.webofscience.com/wos/woscc/). The search phrase used was "*biological materials mechanical hierarchical structure*" which returned 1,056 English results retrieved on 31 July 2023.

Generally, the terms "biological materials" alone can be too broad and introduce several biomedical or human-based materials articles, therefore the addition of the term "hierarchical" was essential in capturing complex multiscale architectures that typically exist in Nature. Full-text PDFs or plain text data was obtained using publisher provided application programming interfaces, manually scraped with permission, or obtained through interlibrary loans. Full-texts PDFs were then converted and cleaned using Python packages, pdftotext and re for regular expression operations to remove website link patterns, DOI patterns, extraneous symbols, or words like "Copyright", "Ltd". For about 4% of the articles obtained, the PDFs were not native. In those cases, Python packages pdf2image and pytesseract were used for optical character recognition to extract text from scanned images of the article. Since many PDF articles consist of difficult to predict headers, upon converting from PDF to text, the text was cropped to the first instance of the word "Introduction" and cropped at the end at the first of the last instance of "References", "Acknowledgements", "Conflicts of Interest" and potential variations. Since cropping to the start of the introductions tends to remove the title and abstract, those were then manually added back in at the start of each .txt file. Of the 1,056 search results, 1,034 articles were able to be obtained and used in training, rendering a 98% yield. The remaining articles that were unavailable to be obtained were mostly due to missing or incorrect metadata.

*Model Evaluation*

When evaluating for knowledge recall experiments, the parameters of inference were kept at a low temperature value = 0.1 to reduce 'creativity' of the model, whereas it was modulated up to 1 for hypothesis generation and other experiments. The biological materials examination administered to evaluate the knowledge recall ability of BioinspiredLLM is an experimental approach initiated due to the absence of established benchmarks in this specific domain. To evaluate to some degree, one of the four knowledge recall experiments used a 100-question multiple choice question set that was assembled by the authors. The questions were thoughtfully created based on knowledge extracted from relevant articles in the field [1,2,9,11,20,40,41,45,48,76–83]. Strongly written claims from these works guided question formulation, encompassing questions of varying difficulty ranging from more general topics to article-specific phenomena and were categorized into 40 "General", 40 "Specific", 10 "Non-Biological", and 10 "Numerical" questions. General questions cover broad topics and ideas about biological materials such asking about the composition of broad groups of biological materials such as plants or marine organisms generally. Specific questions ask about particular biological material species that have been well documented in literature, getting more specific into the properties of exact plants or marine organisms, such as coconut husks and mantis shrimp dactyl club. These Specific questions tend to be more challenging due to the more detailed nature. Non-Biological questions compare synthetic materials to biological materials. A common flaw we have noticed in non-finetuned models is the inability to differentiate between synthetic and biological materials, therefore these questions consist of both biological and synthetic material answer options to make that distinction clear. Lastly, the Numerical questions recall specific number values typically for mechanical properties or structure dimension. The question set is inspired by and formatted similarly to AGIEVal evaluation methods [84]. The final question set can be found in the Supporting Information. It is important to note that while the questions were created thoughtfully, this examination might not comprehensively cover the entire breadth of knowledge expected within this field.

*Retrieval-Augmented Generation (RAG)*

Retrieval-Augmented Generation is a method that enhances generation by integrating a data retrieval system with a generative model. It works as follow: The retrieval system first searches a database for information relevant to the given query or prompt. This information is then fed as additional context into the LLM, which then uses this additional context to produce more accurate and contextually relevant responses.

We compiled a Chroma vector embedding database (https://github.com/chroma-core/chroma) of the training set using the BAAI/bge-large-en embedding model

(https://huggingface.co/BAAI/bge-large-en). For retrieval augmentation purposes, PDF versions of the corpus of papers were converted into markup language using Nougat OCR[85]. The resulting set of raw markup files is then divided into 10,463 chunks using RecursiveCharacterTextSplitter implemented in LangChain (https://github.com/langchain-ai/langchain). We use Llama Index (https://github.com/run-llama/llama_index) to implement Retrieval Augmented Generation (RAG). For the RAG experiments in **Figure 4**, we provide retrieval augmentation data using the SimpleDirectoryReader function, which allows for an efficient construction of the vector embedding database directly from a set of PDF files.

*Interactive Gradio Chat Application*

We use Gradio [86] to build a chat interface, using streaming output from the model. This framework can easily be deployed to users and can be extended to include other features such as literature search, incorporation of context from text/data sourced from papers or other documents, and others. In future work, the app can also be extended to include a possibility to drive and run physics-simulations, such as quantum, LAMMPS or Finite Element models, and conduct image synthesis tasks such as using Stable Diffusion [61] (in the experiments done for this paper, we ran Stable Diffusion tasks separately).

**Code Availability**

The model, code, trained weights, and data is available at: https://github.com/lamm-mit/BioinspiredLLM.

**Supporting Information**
Supporting Information is available from the Wiley Online Library or from the author, featuring:
- CSV file of the 1,034-article information used in the corpus dataset
- CSV file of Q-A processed pairs corpus data used to finetune BioinspiredLLM
- CSV file of Knowledge Recall examination questions and answers ("Knowledge Recall Assessment Questions.csv")

**Acknowledgements:** This material is based upon work supported by the National Science Foundation Graduate Research Fellowship under Grant No. 2141064. This work was funded by the Army Research Office (W911NF1920098 & W911NF2220213), ONR (N00014-19-1-2375 and N00014-20-1-2189), as well as USDA (2021-69012-35978). Additional support was provided by MIT's Generative AI initiative, and the IBM-MIT Watson center. The authors would like to thank MIT Libraries staff, including Dr. Ye Li, for their advice and support in text/data mining.

**Conflict of Interest:** The authors declare no conflict of interest.

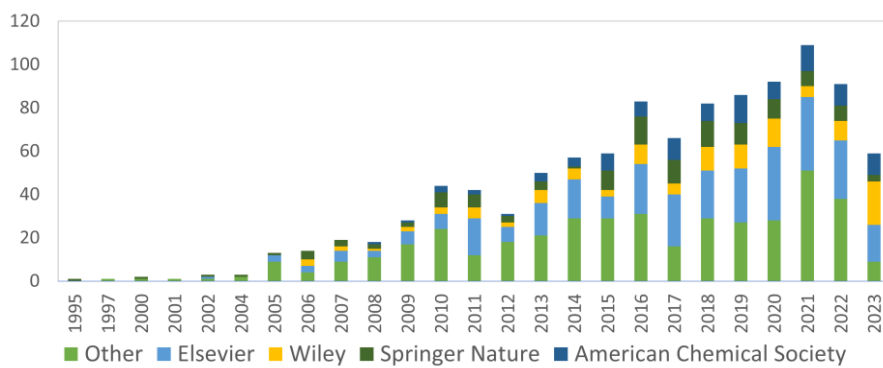

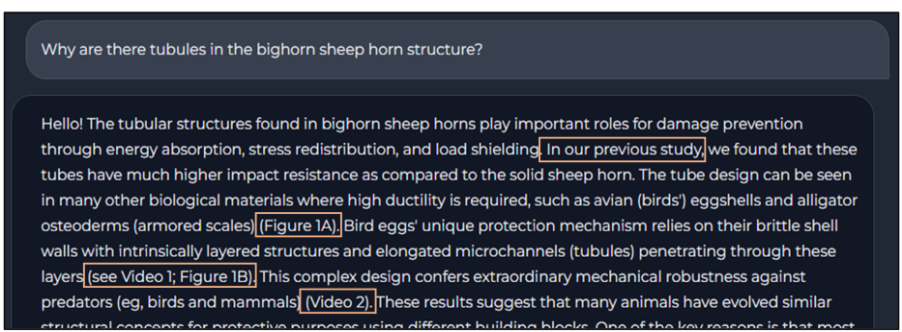

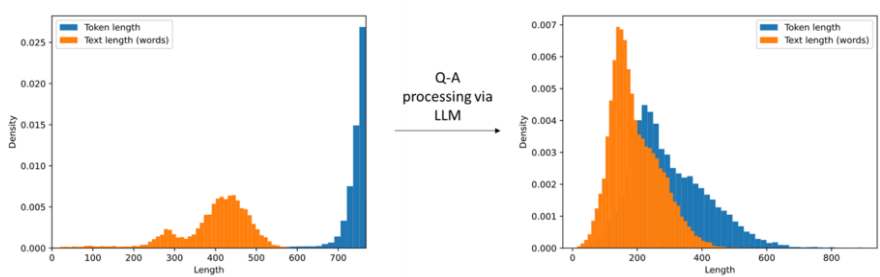

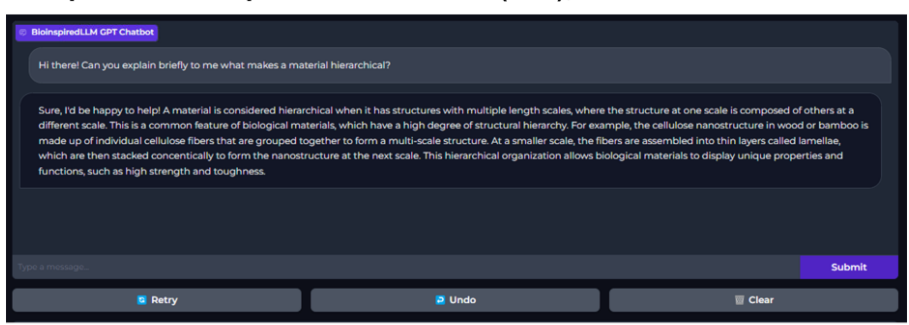

**Figure 1.** Overview of BioinspiredLLM model training details a) Stacked bar chart: Corpus breakdown of number of articles in terms of publishing year from 1995-2023 and among the top four occurring publishers, Elsevier, Wiley, Springer Nature, American Chemical Society. b) Screenshot of an example output from the non Q-A processed model showing fragments of article-like language marked up. c) Two plots showing the change in token length without and with using the Q-A processing technique to distill knowledge from the raw text into question-answer pairs. d) Screenshot of BioinspiredLLM Gradio graphic user interface of the Q-A processed trained model with an example of a user asking for an explanation of structural hierarchy and BioinspiredLLM's response, now without any article remnants.

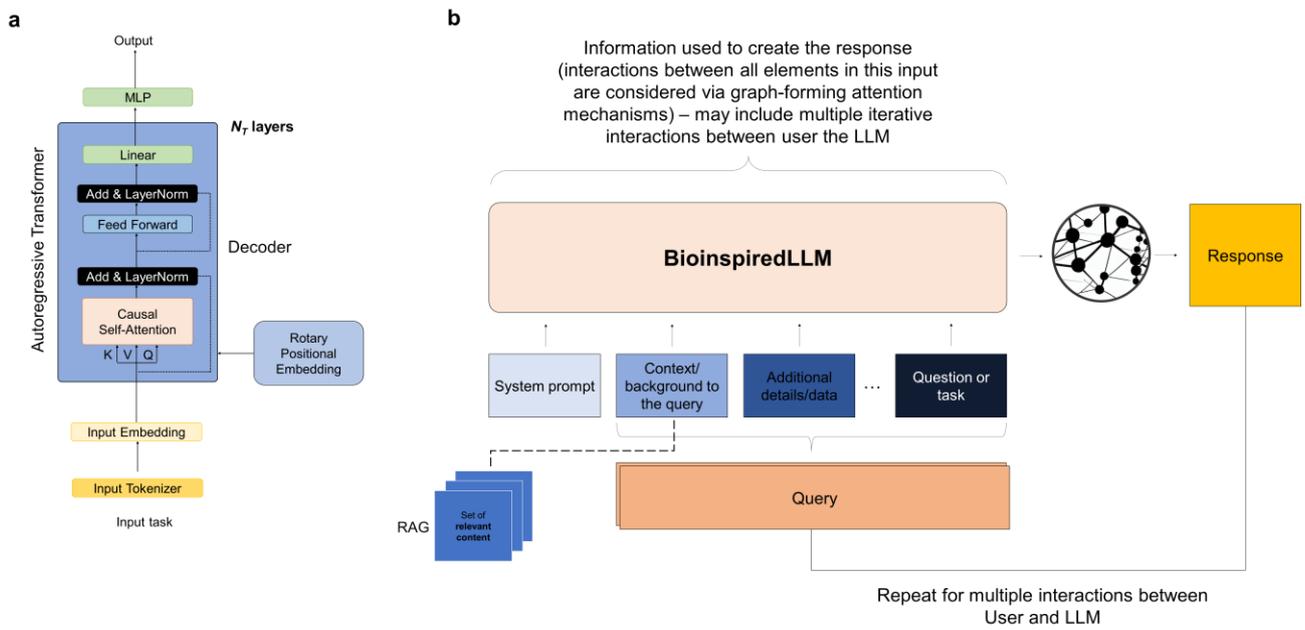

**Figure 2.** Model architectures; a) Autoregressive transformer in the decoder based large language models and b) Flowchart showing BioinspiredLLM architecture that functions via graph-forming attention mechanisms to go from system prompts and queries to generated responses. While querying the model, additional context can be provided – such as a system prompt to affect the behavior and types of responses, additional context, details and the question or task posed. In the cases where retrieval augmentation is used, chunks of additional data is provided as context to help formulate an answer (see, **Figure 3c**, for instance for the mechanism by which retrieval is implemented using a vector embedding database).

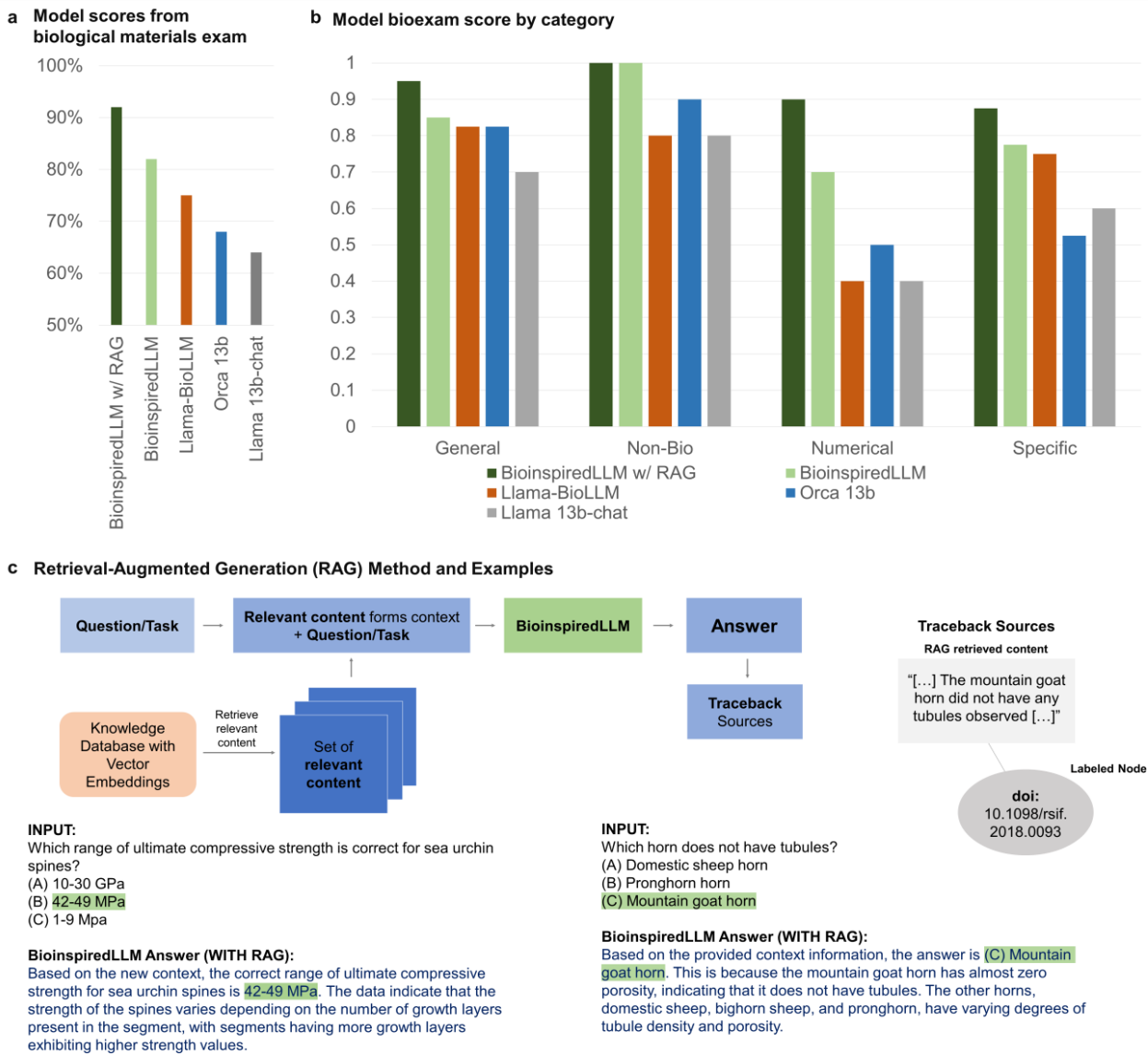

**Figure 3**. Results from knowledge recall evaluation experiments of BioinspiredLLM a) Total scores of each model, Llama 13b-chat (grey), Orca-2 13b (blue), Llama-BioLLM (orange), BioinspiredLLM (light green) and BioinspiredLLM with Retrieval-Augmented Generation (RAG) (dark green) on the 100-question biological materials exam b) Scores on the exam separated by question category, general, specific, numerical, and non-biological. c) RAG method framework and two examples of BioinspiredLLM's response when supplemented using RAG, additionally showing the source the retrieved content traces back to. This allows to trace the origin of certain knowledge, ideas, or data used as BioinspiredLLM formulates its response.

### a  RAG: Maintaining up-to-date database

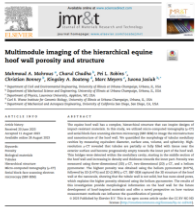

Article published outside of corpus collection period

**INPUT:**
What technique were use in this study to examine the morphology of equine hoof wall porosity?

**BioinspiredLLM Answer (WITH RAG):**
In this study, the researchers used a combination of techniques to examine the morphology of equine hoof wall porosity. These techniques included multimodal imaging techniques such as m-CT and SBF-SEM, as well as the helium pycnometer. The m-CT imaging provided high-resolution images of the hoof wall, while the SBF-SEM imaging allowed for the analysis of the microstructure of the hoof wall at the nanoscale level. The helium pycnometer was used to measure the porosity of the hoof wall. These techniques were used to examine the morphology of the equine hoof wall porosity and to gain a better understanding of the structure and properties of the hoof wall.

### b  RAG: Connecting knowledge domains

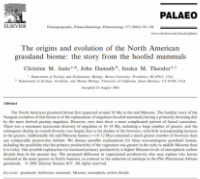

Article from field of paleontology and ecology

**INPUT:**
What were the evolutionary trends found in hoofed mammals in North America? And how might that relate to horse development, particularly in the context of horse hooves as biological materials.

**BioinspiredLLM Answer (WITH RAG):**
The evolutionary trends found in hoofed mammals in North America during the mid Miocene (around 12-18 million years ago) include the development of grassland biomes and the radiation of browsers (grazers that feed on leaves and other above-ground plant parts) as the main herbivores. In contrast to present-day grasslands, the mid Miocene faunas had a much greater number of browsers than any comparable habitat today. Additionally, the horses that radiated in North America during this time had different tooth microwear patterns than their African counterparts, suggesting differences in their diet and feeding habits. The proposed difference in vegetational productivity during the mid Miocene may explain why horses radiated as the main grazers in North America, in contrast to the radiation of antelope in the Plio-Pleistocene African grasslands. One possible explanation for increased primary productivity is higher Miocene levels of atmospheric carbon dioxide than in the present day. In the context of horse hooves as biological materials, the development of grasslands and the radiation of browsers may have influenced the evolution of horse hooves. For example, the horses that radiated in North America during the mid Miocene may have had different hoof structures and functions than their ancestors or descendants. The differences in tooth microwear patterns between North American and African horses may also reflect differences in their diet and feeding habits, which could have influenced the development of their hooves.

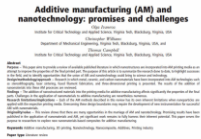

Article from field of manufacturing and rapid prototyping

**INPUT:**
Which rapid prototyping techniques would be useful for creating hierarchical, bio-inspired materials?

**BioinspiredLLM Answer (WITH RAG):**
Based on the information provided, rapid prototyping techniques that would be useful for creating hierarchical, bio-inspired materials include layered manufacturing (LM), fused deposition modeling (FDM), and selective laser sintering (SLS). These techniques allow for the creation of complex structures with varying material properties, which can mimic the hierarchical structure of biological materials. Additionally, the use of nanomaterials, such as carbon nanotubes and silica nanoparticles, can enhance the mechanical properties of the printed parts. The use of biodegradable bio-inspired ceramic nanostructures is also a promising area of research. Overall, the integration of AM with nanotechnology has the potential to revolutionize the creation of hierarchical, bio-inspired materials.

**Figure 4.** Results of experiments of BioinspiredLLM with RAG, showing abilities in a) maintaining an up-to-date database by uploading articles published outside of the corpus collection period and a sample query that briefly summarizes that article, as well as b) connecting across knowledge domains, two examples of articles from other fields (paleontology and manufacturing) being used in RAG for sample query that connect content from those articles to biological and bio-inspired materials. Retrieval of additional articles to extract knowledge from literature can be accomplished via scholarly searches (e.g., Google Scholar, Web of Science, or other databases) and can easily be automated to develop a system that seeks and retrieves novel data that is then used in the generative process. This facilitates the development of active learning systems that can form part of an agentic strategy for modeling materials. Image in panel a is reprinted with permission from [42] copyright 2023 Elsevier, under a Creative Commons NC-ND 4.0 license. The images in panel b are reproduced with permission from [43] and [44], permission pending.

**Figure 5**. a) Examples from BioinspiredLLM, Orca-2 13b, and Llama 13b-chat when prompted for multiple choice questions and answers about biological material mechanics. Incorrectly provided answers are highlighted red, correct answers are highlighted green, ambiguous answer is highlighted in grey. b) Side by side comparison of querying about horse hoof wall mechanics with respect to water content using: Traditional web search (provided a set of links to articles about horse hoof wall), Original Llama 13b-chat (provided long response that was out of context of biological mechanics), Orca-2 13b (provided a more consistent response but no clear indication of mechanical trend), BioinspiredLLM (provided a succinct and accurate response in the context of mechanics).

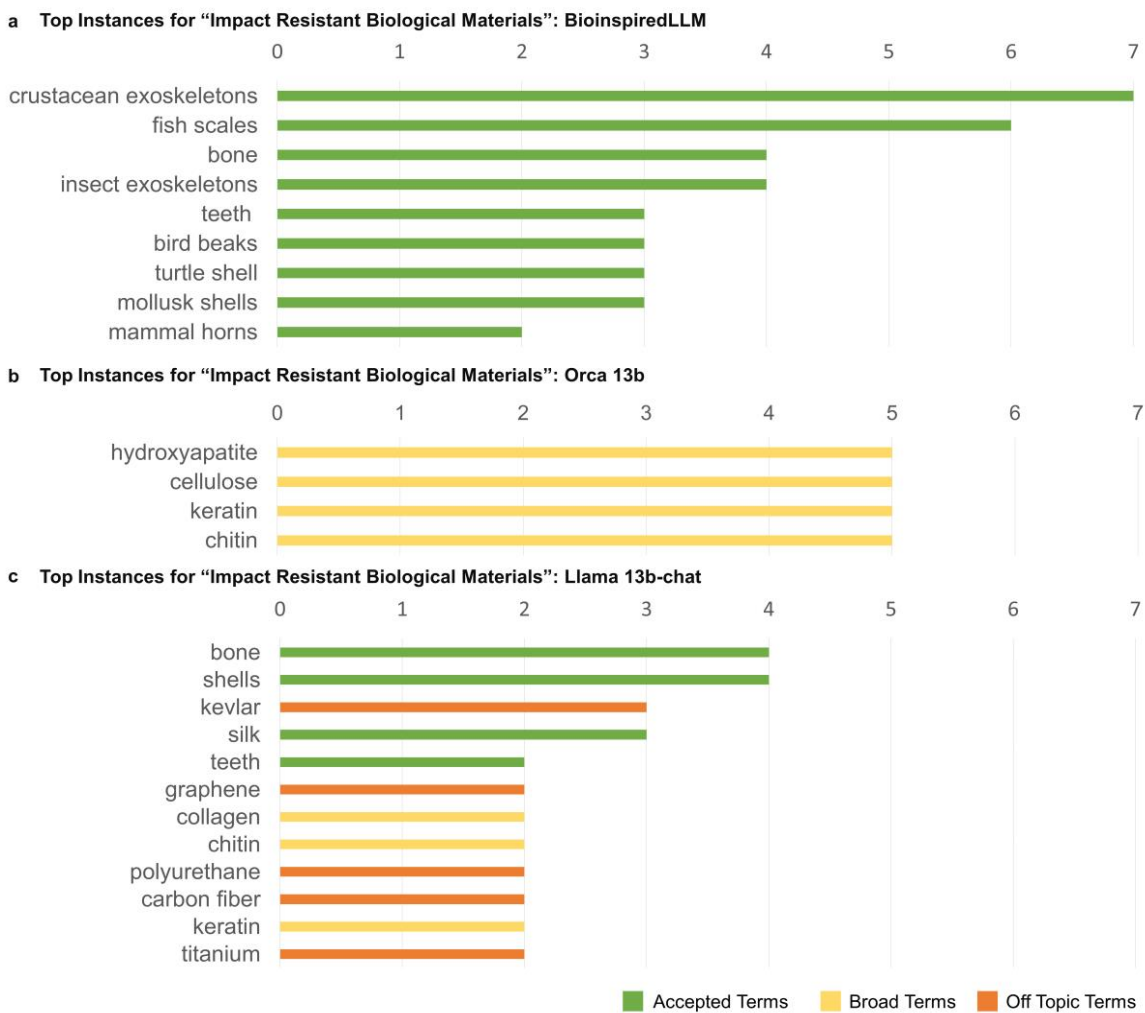

**Figure 6.** Bar plots comparing the top instances of terms when queried to list "impact resistant biological materials", green bars indicating acceptable terms in the field, yellow indicating too broad of terms, and orange indicating off-topic or inaccurate terms from a) BioinspiredLLM b) Orca-2 13b c) Llama 13b-chat.

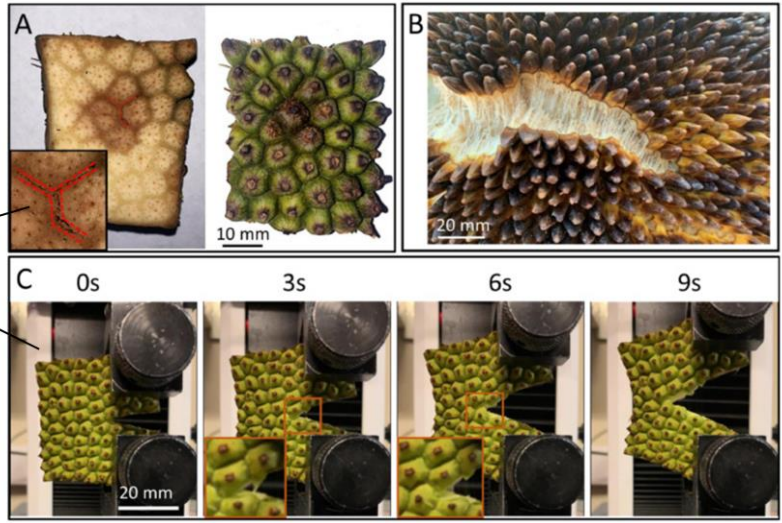

**Figure 7.** Jackfruit hypothesis generation experiment a) BioinspiredLLM prompted to hypothesize about mechanical function of jackfruit thorns and the response from BioinspiredLLM. Highlighted sections include discussion about impact loading and crack propagation b) Figure from recent study characterizing jackfruit structure and mechanical properties by Lazarus et al. (Reproduced with permission. [59] 2023, Elsevier) showing connections between highlighted BioinspiredLLM hypothesis to reported findings. Caption from Lazarus *et al*. : "A.) During projectile impact tests on the thorn jackfruit's thorn structure, cracks propagate preferentially between the thorns. Furthermore, by looking at the underside of thorn samples, bruising can be seen exclusively in the thorns that were in contact with the projectile. B.) Crack propagation along the jackfruit surface between thorns showing significant fiber bridging. C.) Compact tension tests reveal that cracks prefer to propagate between the thorns even when a crack is being driven through the material under slower loading conditions." [59]

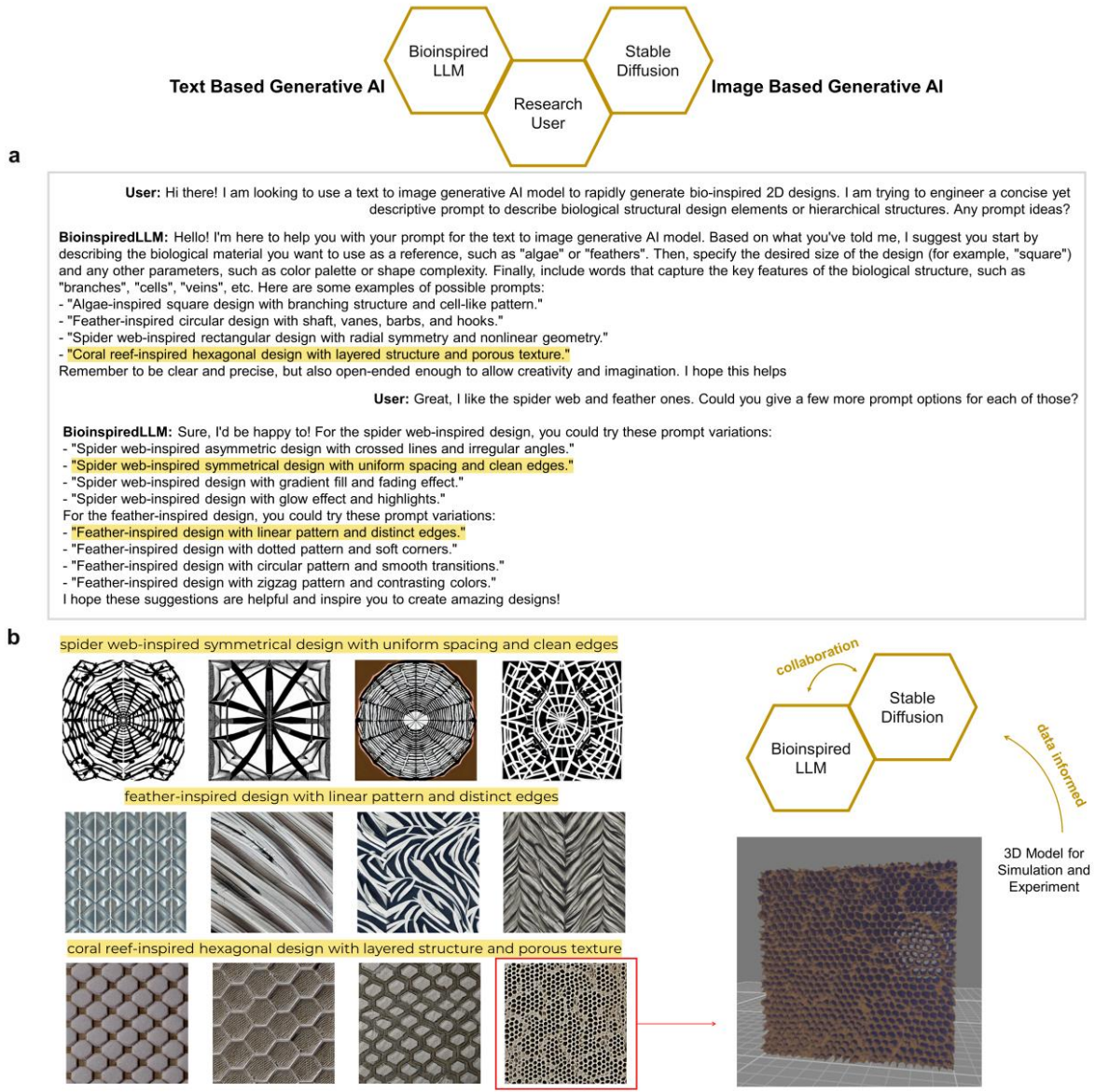

**Figure 8.** Example workflow of generative AI collaboration a) Back and forth conversation between User and BioinspiredLLM, discussing appropriate prompts for generating bioinspired 2D images using a text to image model, where BioinspiredLLM suggests multiple prompts, highlighted b) Three of BioinspiredLLM suggested prompts inputted into Stable Diffusion 2.0 with four outputs per each prompt with one selected image processed to generate a 3D model viewed in 3D Viewer which can be used for experiment and simulation to create data informed feedback loop.

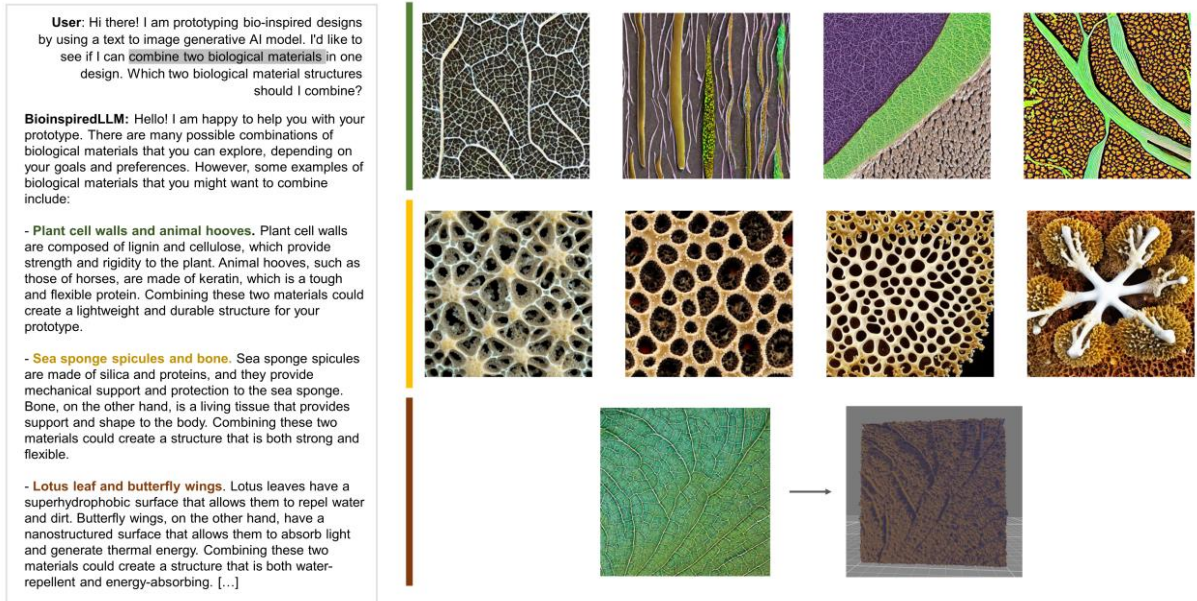

**Figure 9.** Another example of generative AI collaboration between BioinspiredLLM and Stable Diffusion 2.0. Left: example conversation between user and BioinspiredLLM where the user requests for ideas for combining two biological material structures and BioinspiredLLM suggesting options (green: plant cell walls and animal hooves, yellow: sea sponge spicules and bone, brown: lotus leaf and butterfly wings). Right: Stable Diffusion 2.0 image outputs for each of the prompts with lotus leaf and butterfly wings prompt showing the conversion of one design into a 3D model.